\documentclass[aps,pra,groupedaddress,showpacs]{revtex4}
\begin{document}

%\preprint{}

\title{A strongly perturbed quantum system is a semiclassical system}


\author{Marco Frasca}
\email[]{marcofrasca@mclink.it}
%\thanks{}
\affiliation{Via Erasmo Gattamelata, 3 \\ 00176 Roma (Italy)}


\date{\today}

\begin{abstract}
%We show that a strongly perturbed quantum system is just a semiclassical system
%being characterized by the Wigner-Kirkwood expansion for the propagator with the same expansion
%for the eigenvalues as for the WKB series. The series for a strong perturbation
%is proved to be dual to the standard weak perturbation theory. 
We show that a strongly perturbed quantum system, being a semiclassical system
characterized by the Wigner-Kirkwood expansion for the propagator, has the same expansion
for the eigenvalues as for the WKB series. The perturbation series is rederived by the
duality principle in perturbation theory.
\end{abstract}

%\keywords{}

\maketitle


Recently we proposed a strongly coupled quantum field theory \cite{fra1}. This approach is
based on the duality principle in perturbation theory as exposed in \cite{fra2} and
applied in quantum mechanics. In this latter paper we showed that the dual perturbation
series to the Dyson series, that applies to weak perturbations, is obtained through
the adiabatic expansion when reinterpreted as done in \cite{most}. This expansion grants
a strong coupling expansion having a development parameter formally being the inverse
of the parameter of the Dyson series and then holding just in the opposite limit
of the perturbation going to infinity.

Due to the relevance of this approach for quantum field theory, it is of paramount
importance to show how this method is able to perform with an anharmonic oscillator as
this example has been set as a fundamental playground by Bender and Wu \cite{bw} in
their pioneering works. Besides, it would be helpful to trace the way the eigenvalues
are obtained in this case in view of methods, as the one of Janke and Kleinert\cite{jk},
providing excellent computational tools to this aims.

Indeed, in this paper we show that our method provides a sound solution to the strong coupling
computation of the ground state energy of the anharmonic oscillator, being this given
by the well known semiclassical series for the eigenvalues in the WKB approximation. The
relevance of the result relies on the fact that we obtain this result from the Wigner-Kirkwood (WK)
expansion that our method yields. This provides an important conceptual result as, besides
showing that the two approaches WKB and WK are equivalent in the strong coupling limit, 
% Modified 7 july 2006
we will be able to rederive the fact that \cite{sim} {\sl a strongly perturbed quantum system is a
semiclassical system} from the duality principle in perturbation theory granting in this
way the soundness of the approach. 
% end modification
This fact may be interesting e.g. in the problem of the measure in
quantum mechanics as we already pointed out in \cite{fra3,fra4} for QED.

The result is obtained by exploiting the link between the Wigner-Kirkwood expansion and
the Thomas-Fermi approximation \cite{rs}. So, one has that a semiclassical approximation
relies at the foundation of a many-body approach and is equivalent to the WKB approximation
in the strong coupling limit giving the same results for the eigenvalues. Indeed, the
Thomas-Fermi approximation is a semiclassical approximation too and this fact is well-known
having been proved by Lieb and Simon \cite{ls1,ls2}. The interesting aspect here is that
such a many-body approach gives back the Bohr-Sommerfeld quantization rule and we
obtain also the higher order corrections.

In order to start our proof, we consider the simple case of a free particle undergoing the
effect of a perturbation $V(x)$ in a one dimensional setting. So, the Schr\"odinger
equation for the propagator $U$ can be written down as
\begin{equation}
    -\frac{\hbar^2}{2m}\frac{\partial^2 U}{\partial x^2}+\lambda V(x)U -i\hbar\frac{\partial U}{\partial t}=0
\end{equation}
being $\lambda$ an ordering parameter and $U(t_0,t_0)=I$. 
The Dyson series for the propagator can straightforwardly be
obtained when the limit $\lambda\rightarrow 0$ is considered giving back the well known solution series
\begin{eqnarray}
      U(t,t_0) &=& U_0(t,t_0)\left[I-\frac{i}{\hbar}\lambda\int_{t_0}^t dt'U_0^{-1}(t',t_0)V(x)U_0(t',t_0)\right. \\ \nonumber
      &-& \left.\frac{\lambda^2}{\hbar^2}\int_{t_0}^t dt' \int_{t_0}^{t'} dt''
	    U_0^{-1}(t',t_0)V(x)U_0(t',t_0)U_0^{-1}(t'',t_0)V(x)U_0(t'',t_0)+\ldots\right]
\end{eqnarray}
being $U_0$ the solution of the equation
\begin{equation}
    -\frac{\hbar^2}{2m}\frac{\partial^2 U_0}{\partial x^2}-i\hbar\frac{\partial U_0}{\partial t}=0
\end{equation}
given by
\begin{equation}
     U_0(t,t_0)=
     \left[\frac{m}{2\pi i\hbar(t-t_0)}\right]^\frac{1}{2}e^{i\frac{m(x-x')^2}{2\hbar(t-t_0)}}.
\end{equation}
We recognize the interaction picture working here.
At this point it is interesting to note that the choice of a perturbation is completely arbitrary and one
may ask what meaning could be attached to a series, with $\lambda=1$, 
where we take $-\frac{\hbar^2}{2m}\frac{\partial^2}{\partial x^2}$ instead of $V(x)$ as a 
perturbation giving the series
\begin{eqnarray}
\label{eq:K}
      K(t,t_0) &=& K_0(t,t_0)\left[I+\frac{i}{\hbar}\int_{t_0}^t dt'
	    K_0^{-1}(t',t_0)\frac{\hbar^2}{2m}\frac{\partial^2}{\partial x^2}K_0(t',t_0)\right. \\ \nonumber
      &-& \frac{1}{\hbar^2}\left.\int_{t_0}^t dt' \int_{t_0}^{t'} dt''
      K_0^{-1}(t',t_0)\frac{\hbar^2}{2m}\frac{\partial^2}{\partial x^2}K_0(t',t_0)
      K_0^{-1}(t'',t_0)\frac{\hbar^2}{2m}\frac{\partial^2}{\partial x^2}K_0(t'',t_0)+\ldots\right]
\end{eqnarray}
being now $K_0$ the solution of the equation
\begin{equation}
    V(x)K_0 -i\hbar\frac{\partial K_0}{\partial t}=0
\end{equation}
being
\begin{equation}
\label{eq:K0}
    K_0(t,t_0) = e^{-\frac{i}{\hbar}V(x)(t-t_0)}.
\end{equation}
The answer can be immediately obtained when, after reinserting the ordering parameter $\lambda$,
we recognize that with a rescaling of time $t\rightarrow\lambda t$ and taking the series
\begin{equation}
    K = K_0 +\frac{1}{\lambda}K_1+\frac{1}{\lambda^2}K_2 + \ldots
\end{equation}
we recover the series (\ref{eq:K}) that is a strong coupling expansion. Setting $\tau=\lambda t$ we
can write it as
\begin{eqnarray}
\label{eq:K1}
      K(\tau,\tau_0) &=& K_0(\tau,\tau_0)\left[I+\frac{i}{\hbar\lambda}\int_{\tau_0}^{\tau} d\tau'
	    K_0^{-1}(\tau',\tau_0)\frac{\hbar^2}{2m}\frac{\partial^2}{\partial x^2}K_0(\tau',\tau_0)\right. \\ \nonumber
      &-& \left.\frac{1}{\hbar^2\lambda^2}\int_{\tau_0}^{\tau} d\tau' \int_{\tau_0}^{\tau'} d\tau''
      K_0^{-1}(\tau',\tau_0)\frac{\hbar^2}{2m}\frac{\partial^2}{\partial x^2}K_0(\tau',\tau_0)
      K_0^{-1}(\tau'',\tau_0)\frac{\hbar^2}{2m}\frac{\partial^2}{\partial x^2}K_0(\tau'',\tau_0)+\ldots\right]
\end{eqnarray}
We easily recognize that this series
is dual to the Dyson series in the sense of Ref.\cite{fra2} having a development parameter that is formally the 
inverse with respect to that of the Dyson series. For this reason we call 
this dual representation the free picture.
This series is the 1+0 dimensional solution of the quantum field
theory we presented in Ref.\cite{fra1} and our aim is to get from this the ground state energy showing that, in
this case, we are working with a semiclassical expansion.

For our aims, we need to exploit this strong coupling series to unveil its nature. We already see from the leading order
solution (\ref{eq:K0}) that it coincides with the leading order of the well-known semiclassical Wigner-Kirkwood
series \cite{rs,sch,jk} 
% Modified 7 july 2006
as should be expected \cite{sim}. 
We need to prove this also for higher orders. Indeed, the computation is straigthforward and
gives, at least for the first correction,
\begin{eqnarray}
\label{eq:WK}
      K(\tau,0) &=& K_0(\tau,0)\left\{I-\frac{1}{\lambda}\left[\frac{i\tau^3}{6m\hbar}(\partial_xV)^2-
      \tau^2\left(\frac{1}{4m}\partial_x^2V+
      \frac{i}{2m\hbar}\partial_xVp\right)+\frac{i}{\hbar}\frac{p^2}{2m}\tau\right]+\ldots
      \right\}
\end{eqnarray}
that is what one should expect for the WK expansion. 
So, we have obtained the semiclassical Wigner-Kirkwodd series out of a dual strong coupling
expansion for a quantum mechanical system by the duality principle in perturbation theory. 
This in turn means, 
% Modified 7 july 2006
as already known \cite{sim}, 
% end modification
that a strongly perturbed quantum system
is a semiclassical system. 
% Modified 7 july 2006
This is in agreement with the fact that a large mass expansion in quantum mechanics
gives rise to a WK expansion out of the WKB expansion\cite{om}. 
% end modification
We just point out that a wide application of the WK expansion is seen in statistical
mechanics \cite{sch} and in this case is obtained with the standard substitution $t\rightarrow -i\hbar\beta$
that we will use in the following.

The WK series appears rather singular depending on ascending power of $\tau$ and having terms proportional to
gradients of the potential making the dependence on $\lambda$ anomalous at best for such an expansion. But as
we will se below, the series for the ground state energy is
well defined and in agreement with the corresponding expression for a WKB series in spite of the very singular
nature of this expansion. This is easy to prove using techniques of many-body physics that will permit us
to show that the leading order of the WK expansion is the well known Thomas-Fermi approximation.

The next step is to recognize that we can resum all the terms with $p^2/2m$ giving finally, after
projecting on the momentum eigenstates with a Wigner transformation of the propagator,
\begin{eqnarray}
\label{eq:WK2}
      C(\beta) &=& C_0(\beta)\left\{I-
      \hbar^2\lambda\beta^2\left(\frac{1}{4m}\partial_x^2V+\frac{i}{2m\hbar}\partial_xVp\right)
	  +\frac{\hbar^2\lambda^2\beta^3}{6m}(\partial_xV)^2
      +\ldots
      \right\}
\end{eqnarray}
being now
\begin{equation}
     C_0(\beta) = e^{-[\frac{p^2}{2m}+\lambda V(x)]\beta}.
\end{equation}
The density matrix can be obtained by inverse Laplace transforming the series (\ref{eq:WK2}) divided
by $\beta$ \cite{rs}, that is
\begin{equation} 
    \rho(x,p,E)=\frac{1}{2\pi i}\int_{c-i\infty}^{c+i\infty}e^{\beta E}\frac{C(\beta)}{\beta}d\beta,
\end{equation}
with $c>0$, giving the expression
\begin{eqnarray}
\label{eq:rho}
    \rho(x,p,E)&=&\theta\left(E-\frac{p^2}{2m}-\lambda V(x)\right)
	  -\lambda\left(\frac{\hbar^2}{4m}\partial_x^2V+
      \frac{i\hbar}{2m}\partial_xVp\right)
	  \delta'\left(E-\frac{p^2}{2m}-\lambda V(x)\right) \\ \nonumber
	 &+&\frac{\hbar^2\lambda^2}{6m}(\partial_xV)^2
	\delta''\left(E-\frac{p^2}{2m}-\lambda V(x)\right)+\ldots
\end{eqnarray}
and the leading order is just the Thomas-Fermi approximation. This series has not a clear dependence on $\lambda$
but this is a strong coupling series that has meaning in the sense of distributions.

We now impose the normalization condition
\begin{equation}
    \int\frac{dpdx}{2\pi\hbar}\rho(x,p,E) = n+\frac{1}{2}
\end{equation}
being $n=0,1,2,\ldots$,
and this gives the WKB energy levels and their higher orders corrections as we will see in a while. 
Indeed, we can substitute eq.(\ref{eq:rho})
into the above normalization condition giving the series \cite{bow}
\begin{eqnarray}
\label{eq:E}
    & &\frac{1}{\pi\hbar}\int_{x_{T1}}^{x_{T2}}dx\sqrt{2m(E-V(x))}-
    \lambda\frac{\hbar}{4\pi}\frac{d}{dE}\int_{x_{T1}}^{x_{T2}}dx\frac{\partial_x^2V(x)}{\sqrt{2m(E-V(x))}} \\ \nonumber
    &+&\lambda^2\frac{\hbar}{6\pi}\frac{d^2}{dE^2}\int_{x_{T1}}^{x_{T2}}dx\frac{(\partial_xV(x))^2}{\sqrt{2m(E-V(x))}}
    +\ldots=n+\frac{1}{2}
\end{eqnarray}
being $x_{T1}$ and $x_{T2}$ the solutions of the equation $E=\lambda V(x)$ that determine the region where the integral 
is meaningful. Use has been made of the fact that $\int\partial_xVp\delta'\left(E-\frac{p^2}{2m}-\lambda V(x)\right)dpdx=0$.
We recognize here the WKB quantization condition and its higher order corrections as promised.

Our aim now is to apply the above expansion to the computation of the ground state energy of a pure quartic
oscillator $H=p^2/2m+\lambda x^4/4$. From \cite{jk} we know that at the leading order one should have
$E=0.667986259155777\ldots(\lambda/4)^\frac{1}{3}$. We know also from \cite{bow} that this value can be
obtained by pushing the series (\ref{eq:E}) to the higher orders. In order to have an idea of what we get,
for our example we get the series
\begin{equation}
     c_0\frac{\sqrt{2}}{\pi}\tilde E_n-\left(c_1\frac{3}{4\sqrt{2}\pi}-c_2\frac{5}{6\sqrt{2}\pi}\right)
     \frac{1}{\tilde E_n}+\ldots = n+\frac{1}{2}
\end{equation}
being
\begin{eqnarray}
    c_0&=&\frac{2\sqrt{2}}{3}K\left(\frac{\sqrt{2}}{2}\right) \\ \nonumber
    c_1&=&-\sqrt{2}K\left(\frac{\sqrt{2}}{2}\right)+2\sqrt{2}E\left(\frac{\sqrt{2}}{2}\right) \\ \nonumber
    c_2&=&-\frac{3}{5}\sqrt{2}K\left(\frac{\sqrt{2}}{2}\right)+\frac{6}{5}\sqrt{2}E\left(\frac{\sqrt{2}}{2}\right)
\end{eqnarray}
being $K$ and $E$ the elliptic integrals of first and second kind respectively, and $\tilde E_n =\left(E_n/(\lambda/4)^{\frac{1}{3}}\right)^{\frac{3}{4}}$. The numerical solution is rather satisfactory as
we get from the Bohr-Sommerfeld term that $\tilde E_0=0.5462673253$ that, as is well known, has an error of about 
20\%, while with the first order correction one has $\tilde E_0=0.7496932075$ that improves to 10\% already at
this order. But, as already shown in \cite{bow}, we know that the semiclassical approach is able to produce
the exact value by going to higher orders.

% Modified 7 july 2006
So, we can conclude that 
%a strongly perturbed quantum system is just a semiclassical system being properly
%described by a gradient expansion in the strongly coupling limit,
the Wigner-Kirkwood series produces
eigenvalues through the Bohr-Sommerfeld rule and its higher order corrections. Besides, our result
has been obtained by considering the duality principle in perturbation theory and applying it to the Dyson series
rederiving by this means the equivalence between strong coupling perturbation theory and semiclassical expansion,
making sound the main result of this paper.
% end modification

\begin{acknowledgments}
I would like to thank Hagen Kleinert for inviting me to Berlin to discuss with him these questions.
\end{acknowledgments}

\end{document}